\documentclass[12pt]{article}
\usepackage{jheppub}
\usepackage{enumerate}
\usepackage{epsfig}
\usepackage{amsmath} 
\usepackage{capt-of}
\usepackage{wasysym} 
\usepackage{mathrsfs} 
\usepackage{graphicx} 
\usepackage{amsfonts} 
\usepackage{amsbsy} 
\usepackage{amscd}
\usepackage{multirow}
\usepackage{rotating}
\usepackage{slashed, latexsym, verbatim, cancel} 
\usepackage[utf8]{inputenc}
\usepackage {multirow}
\usepackage{color}
\definecolor{myred}{rgb}{0.6,0,0} 
\definecolor{myblue}{rgb}{0,0.2,0.4}
\definecolor{mygreen}{rgb}{0,0.9,0.1}
\definecolor{hc}{rgb}{.9,0.1,0.7}
\definecolor{hcout}{rgb}{.9,0.7,0.9}
\definecolor{Orange}{rgb}{1.,0.65,0.}

\makeatletter

\newcommand{\fmslash}[2][0mu]{%
  \mathchoice
    {\fmsl@sh\displaystyle{#1}{#2}}%
    {\fmsl@sh\textstyle{#1}{#2}}%
    {\fmsl@sh\scriptstyle{#1}{#2}}%
    {\fmsl@sh\scriptscriptstyle{#1}{#2}}}
\newcommand{\fmsl@sh}[3]{%
  \m@th\ooalign{$\hfil#1\mkern#2/\hfil$\crcr$#1#3$}}
\makeatother

\newcommand{\lsim}{{\;\raise0.3ex\hbox{$<$\kern-0.75em\raise-1.1ex\hbox{$\sim$}}\;}}
\newcommand{\gsim}{{\;\raise0.3ex\hbox{$>$\kern-0.75em\raise-1.1ex\hbox{$\sim$}}\;}}

\usepackage{array}
\newcolumntype{C}[1]{>{\centering\arraybackslash$}p{#1}<{$}}

\usepackage{bm} 

\usepackage{cleveref}
\newcommand{\be}{\begin{equation}}
\newcommand{\ee}{\end{equation}}
\newcommand{\bes}{\begin{equation*}}
\newcommand{\ees}{\end{equation*}}
\newcommand{\bea}{\begin{eqnarray}}
\newcommand{\eea}{\end{eqnarray}}
\newcommand{\beas}{\begin{eqnarray*}}
\newcommand{\eeas}{\end{eqnarray*}}

\newcommand{\la}{\lambda}

\newcommand{\wt}{\widetilde}
\title{Searching for a Light Higgs Boson via the Yukawa Process at Lepton Colliders}
\author{Eung Jin Chun,}
\author{Tanmoy Mondal} 
\affiliation{Korea Institute for Advanced Study, Seoul 02455, Korea}
\emailAdd{ejchun@kias.re.kr}
\emailAdd{tanmoy@kias.re.kr}
  
\abstract{
We explore the prospect of Yukawa production of a light boson which can exist in an extended Higgs sector. 
A particularly interesting case is the light pseudoscalar in Type-X two Higgs doublet model which can explain the anomalous magnetic moment of muon at large $\tan\beta$. Considering ILC \emph{Higgs factory} with $\sqrt{s}=250$ GeV, we show that the available parameter space can be fully examined by the  (tau) Yukawa process at $5\sigma$. 
We also demonstrate the mass reconstruction of such a light particle  which helps to sizably minimize the background events.
}

\preprint{KIAS-P19055}
\date{\today}
\begin{document}
\maketitle
\section{Introduction}

Although the Standard Model (SM) endowed with the minimal Higgs sector is enough to explain most experimental data,
it still has room to accommodate an extension with more bosons. In particular, the two Higgs doublet model(2HDM), where two doublet bosons are involved in the electroweak symmetry breaking (EWSB), has been motivated by supersymmetry~\cite{Haber:1984rc}, 
baryon asymmetry of the Universe~\cite{Turok:1990zg,Trodden:1998ym}, or to resolve the strong CP problem~\cite{Kim:1986ax}. Based on the Yukawa structure of natural flavor conservation~\cite{PhysRevD.15.1958}, four different types of 2HDM~\cite{Gunion:1989we,Djouadi:2005gj,Branco:2011iw} can be constructed. 
In the spectrum of multi-Higgs bosons the presence of a relatively light neutral one does not necessarily violate the custodial 
symmetry~\cite{Gerard:2007kn} and thus can be consistent with the electroweak precision test~\cite{Broggio:2014mna}.

A light pseudoscalar in the Lepton-specific or Type-X 2HDM is of particular interest as it can explain the observed anomaly of muon anomalous magnetic moment, ($g-2$)$_\mu$, measured by the BNL collaboration~\cite{Brown:2001mga,Bennett:2006fi} when the ratio of Higgs vacuum expectation values, $\tan\beta$, is sufficiently large~\cite{Cao:2009as,Broggio:2014mna,Ilisie:2015tra,Abe:2015oca,Chun:2016hzs,Cherchiglia:2017uwv,Wang:2018hnw}. 
Such a light pseudoscalar, denoted by $A$, in Type-X 2HDM with large $\tan\beta$ often remains undetected at the large hadron collider (LHC) by virtue of its hadrophbic nature, which makes the production of $A$ via the  gluon fusion process ineffective. 
Possible LHC probes of the parameter space explaining ($g-2$)$_\mu$ include the associated production of $A$ along with a charged ($H^\pm$) or neutral ($H$) scalar~\cite{Chun:2015hsa,Chun:2018vsn} and the decay of the 125 GeV SM Higgs boson ($h$) to a pair of pseudoscalars \cite{Chun:2017yob}.
The associated production can explore the Type-X 2HDM parameter space at high luminosity LHC where the mass of $H^\pm$ or $H$ ($m_{H^\pm},m_H$) is around  200 GeV and $m_A$ is close to 40 GeV~\cite{Chun:2018vsn}.  
On the other hand, the present upper limit on the branching ratio (BR) of the SM Higgs decaying to an $AA$ pair is BR$(h\to A \ A) < 3-4\%$~\cite{Sirunyan:2018mbx}. However, in general the mass of the new scalars ($m_{H^\pm}, m_H$) can be larger than 200 GeV, also the 
$hAA$ coupling is independent of the parameters involving $(g-2)_\mu$ and it is possible that the BR($h\to AA$) can be much smaller than the limit possible to obtain at the LHC~\cite{Chun:2015hsa}, rendering the pseudoscalar virtually untraceable at the LHC. 

\medskip

At a lepton collider  such a light pseudoscalar can be searched  via the Yukawa production channel where a 
light $A$ is radiated from a tau lepton as its coupling to $A$ is large. The Yukawa 
production in Type-II 2HDM was studied~\cite{Kalinowski:1996nr} for LEP-I with a light scalar/pseudoscalar in the 
 $2b\ 2\tau$ final state. There are several proposals for future lepton colliders: ILC~\cite{Baer:2013cma,Bambade:2019fyw}, CEPC \cite{CEPCStudyGroup:2018ghi}, and FCC-ee \cite{Blondel:2019yqr}.  All of these will run as \emph{Higgs factory} where the 
 center-of-mass energy ($\sqrt{s}$) is close to 250 GeV at which the associated production of the SM Higgs boson peaks.
 Hence it is worthwhile to study the prospect of search for a light boson at a 250 GeV lepton collider like ILC which has 
 not been studied before.  In the context of ILC, the Type-X 2HDM model was studied with $\sqrt{s} =$ 500  GeV where the 
 $e^+e^- \to H A \to 4\tau$ is the dominant channel~\cite{Kanemura:2012az}. Similar study was done in 
 $e^+e^- \to H A \to 2 \mu2\tau$ channel at 500 GeV and 1 TeV lepton collider. In both these studies the pseudoscalar 
 is assumed to be heavier than the SM Higgs~\cite{Hashemi:2017awj}.

Here we are interested in the case where $m_{H^\pm}\approx m_H$ is 250 GeV or above, and thus only the Yukawa production of a light boson is feasible at a lepton collider. Focusing on a light $A$ motivated by the $(g-2)_\mu$ measurement, 
we will show how to look for such a particle at a \emph{Higgs factory} in the Yukawa channel with four tau final state. 
The mass of the light pseudoscalar can be 
efficiently reconstructed by using the collinear approximation, and thus it is possible to explore the whole $(g-2)_\mu$ compatible parameter space  of the  Type-X 2HDM at the ILC with 2000 $fb^{-1}$ of integrated luminosity. Note that this search is independent of the  heavy Higgs masses $m_{H^\pm, H}$ and thus directly probes the 2HDM accounting for the $(g-2)_\mu$ anomaly.

\medskip 

The paper is organized as follows, In Section \ref{sec:model}, we give a brief introduction to the Type-X 2HDM along with the 
theoretical and experimental constraints on this model. In Section \ref{sec:simulation}, we describe the methodology of our analysis with 
detail mass reconstruction strategy using collinear approximation and the results of analysis are presented in Section 
\ref{sec:result}. Finally we  conclude in Section \ref{sec:conclusion}. The appendix contains some more details of event simulation.

\section{The Type-X 2HDM}\label{sec:model}

The 2HDM model consists of two scalar doublets $\Phi_1$ and $\Phi_2$ with hypercharge $Y=1$. The model has already been discussed 
in detail elsewhere~\cite{Gunion:1989we,Djouadi:2005gj,Branco:2011iw}. For completeness we will 
briefly discuss the necessary parts of the model, and summarize  various constraints which restricts the model parameters. 

\subsection{Model basics} 

Presence of two Higgs doublets where both the doublets couples to the fermions leads to flavor changing neutral current (FCNC)
interaction at tree level. To avoid this, we can impose the so called Glashow-Weinberg condition  i.e. only one of the two Higgs doublets 
will couple to the right-handed (RH) fermions of the Standard Model~\cite{PhysRevD.15.1958}. This can be realized by imposing an
additional $\mathbb{Z}_2$ symmetry  such that $\Phi_1\rightarrow-\Phi_1$ and $\Phi_2\rightarrow \Phi_2$. The fermions are also charged appropriately under the discrete symmetry. The scalar potential then reads as,
\begin{eqnarray}
\nonumber V_{\mathrm{2HDM}} &=& -m_{11}^2\Phi_1^{\dagger}\Phi_1 - m_{22}^2\Phi_2^{\dagger}\Phi_2 -\Big[m_{12}^2\Phi_1^{\dagger}\Phi_2 + \mathrm{h.c.}\Big]
+\frac{1}{2}\lambda_1\left(\Phi_1^\dagger\Phi_1\right)^2+\frac{1}{2}\lambda_2\left(\Phi_2^\dagger\Phi_2\right)^2 \\
\nonumber && +\lambda_3\left(\Phi_1^\dagger\Phi_1\right)\left(\Phi_2^\dagger\Phi_2\right)+\lambda_4\left(\Phi_1^\dagger\Phi_2\right)\left(\Phi_2^\dagger\Phi_1\right)
+\Big\{ \frac{1}{2}\lambda_5\left(\Phi_1^\dagger\Phi_2\right)^2+  \rm{h.c.}\Big\}.
\label{eq:2hdm-pot}
\end{eqnarray}
The dimensionful coupling $m_{12}^2$ softly breaks the $\mathbb{Z}_2$ charge and for simplicity we have considered that all the 
couplings are real as our results will not depend on it. 
After the electroweak symmetry breaking the scalars $\Phi_1$ and $\Phi_2$ will acquire  vacuum expectation values($vev$) $v_1$ 
and $v_2$ respectively. We can parameterize the doublets in the following way,
$\Phi_j=(H_j^+,(v_j + h_j + i A_j)/\sqrt{2})^T$ and  obtain the five massive physical states $A$ (CP-odd), $h$, $H$, $H^{\pm}$ in terms of
the gauge eigenstates:
\begin{align}
 \begin{pmatrix} H  \\ h \end{pmatrix} =  \begin{pmatrix}  c_{\alpha} && s_{\alpha} \\ -s_{\alpha} &&  c_{\alpha} \end{pmatrix}
  \begin{pmatrix} h_1  \\ h_2 \end{pmatrix}  \label{2hdm_scalar_basis}
 \end{align}
and  $A=-s_\beta \;A_1 + c_\beta \;A_2,\quad H^{\pm}=-s_\beta\; H_1^{\pm} + c_\beta\; H^{\pm}_2$
where $s_\alpha = {\rm sin}~\alpha$, $c_\beta = {\rm cos}~ \beta$ etc and ${\rm tan}~\beta = \cfrac{v_2}{v_1}$ .
The CP-even state $h$ is identified with the SM-like Higgs with mass $m_h \approx 125$ GeV.

Based on the $\mathbb{Z}_2$ charge assignment of the fermions there are  four possible type of Yukawa 
structures and in this article we will consider the lepton specific or Type-X 2HDM where the RH leptons are odd under $\mathbb{Z}_2$ symmetry. The relevant Yukawa Lagrangian is given by,
\begin{equation}\label{eq:yukawa}
-{\cal L}_Y= Y^u\bar{ Q_L} \wt \Phi_2 u_R + Y^d  \bar{ Q_L} \Phi_2 d_R+Y^e\bar{ l_L} \Phi_1 e_R + h.c.,
\end{equation}
where $\wt \Phi_2=i\sigma_2\Phi_2^*$. After symmetry breaking the we can write the Yukawa Lagrangian in terms of mass eigenstates,
\begin{eqnarray}
\nonumber \mathcal L_{\mathrm{Yukawa}}^{\mathrm{Physical}} &=&
-\sum_{f=u,d,\ell} \frac{m_f}{v}\left(\xi_h^f\overline{f}hf +
\xi_H^f\overline{f}Hf - i\xi_A^f\overline{f}\gamma_5Af \right) \\
 &&-\left\{ \frac{\sqrt{2}V_{ud}}
{v}\overline{u}\left(\xi_A^{u} m_{u} P_L+\xi_A^{d} m_{d} P_R\right)H^{+}d  +
\frac{\sqrt{2}m_l}{v}\xi_A^l\overline{v}_LH^{+}l_R + \mathrm{h.c.}\right\},
\label{eq:L2hdm}
\end{eqnarray}
where  $u$, $d$, and $l$ refer to the up-type quarks, down-type quarks, and charged leptons, respectively. 
The Yukawa multiplicative factors, {\it i.e.} $\xi_{\phi}^f$ are given in Table~\ref{Tab:YukawaFactors}. 
 In the limit $\cos(\beta-\alpha) \to 0 $, the modifiers to the SM like Higgs goes to +1 and matches exactly with the SM Yukawa coupling. This is called right sign (RS) Yukawa limit. However, the lepton Yukawa modifier $\xi_{h}^\ell$  becomes `-1' 
 if $\cos(\beta-\alpha)$ is $2/\tan\beta$  and this limit of $(\beta-\alpha)$ is known as wrong sign (WS) Yukawa limit.

\begin{table}[t]
\begin{center}
\begin{tabular}{|c||c|c|c|c|c|c|c|c|c|}
\hline
& $\xi_h^u$ & $\xi_h^d$ & $\xi_h^\ell$
& $\xi_H^u$ & $\xi_H^d$ & $\xi_H^\ell$
& $\xi_A^u$ & $\xi_A^d$ & $\xi_A^\ell$ \\ \hline
Type-X
& $c_\alpha/s_\beta$ & $c_\alpha/s_\beta$ & $-s_\alpha/c_\beta$
& $s_\alpha/s_\beta$ & $s_\alpha/s_\beta$ & $c_\alpha/c_\beta$
& $\cot\beta$ & $-\cot\beta$ & $\tan\beta$ \\
 \hline
\end{tabular}
\end{center}
 \caption{The multiplicative factors of Yukawa interactions in type X 2HDM}
\label{Tab:YukawaFactors}
\end{table}

\subsection{Constraints on the model}

Theoretical constraints on the quartic couplings come from vacuum stability, perturbativity and unitarity. The parameter 
space of Type-X 2HDM has been studied under these conditions and it has been found that~\cite{Broggio:2014mna,Wang:2014sda},
\bea
m_{H} \simeq m_{H^\pm} &\leq& 250 \textrm{ GeV} \hspace{1cm} \textrm{(RS scenario)}\nonumber \\
m_{H} \simeq m_{H^\pm} &\leq& \sqrt{\la_{max}}\ v = \sqrt{4\pi}\ v\hspace{1cm} \textrm{(WS scenario)}.
\eea 
We will appropriately choose the value of $\cos(\beta-\alpha)$ to satisfy these conditions. 

The constraints from the electroweak precision measurements are encoded in S,T and U parameters and it restricts that the charged higgs boson has to be nearly 
degenerate with either $H$ or $A$~\cite{Broggio:2014mna, Haller:2018nnx}. We will assume $m_{H}\simeq m_{H^\pm}$ to satisfy the 
EWPT. The constraints from the searches of extra scalars at the LHC has very little impact on 2HDM-X parameter space as the extra  scalars 
in this model are hadrophobic and their coupling decreases as $\tan\beta$ increases. The limit from LEP on pair production of $A$ 
and $H$ via $Z$ is $m_A+m_H  > 185 $ GeV~\cite{Abdallah:2004wy} which we will respect. Since the new scalars couples to quarks 
very weakly at large  $\tan\beta$ the flavor constrains coming from $B\to X_s\gamma$ or $ B_s\to\mu^+\mu^-$ are extremely weak 
and for $\tan\beta > 5$ there is no limit on the scalar spectrum from hadronic flavor observables~\cite{Haller:2018nnx}. 
The global analysis of the present Higgs 
data was analysed by the GFitter collaboration and the WS limit is allowed for large $\tan\beta$~\cite{Haller:2018nnx}.

We are interested in the parameter space where the pseudoscalar is light $m_A < 90$ GeV with $\tan\beta\gg 1$ as this region can 
explain the muon anomaly. We will assume $m_H=m_{H^\pm}=250$ GeV and wrong sign Yukawa limit which satisfies all the 
theoretical and experimental constraints discussed above as well as limits from precision leptonic observables. 
Due to the leptophilic nature of such a light pseudoscalar, lepton universality tests \cite{ALEPH:2005ab,Amhis:2016xyh} 
can provide severe bounds on the parameter space
favorable for $(g-2)_\mu$ \cite{Abe:2015oca,Chun:2016hzs}.

\section{Search for Yukawa process at lepton collider}\label{sec:simulation}

\begin{figure}[!t]
\begin{center}
 \includegraphics[width=7cm,angle=270]{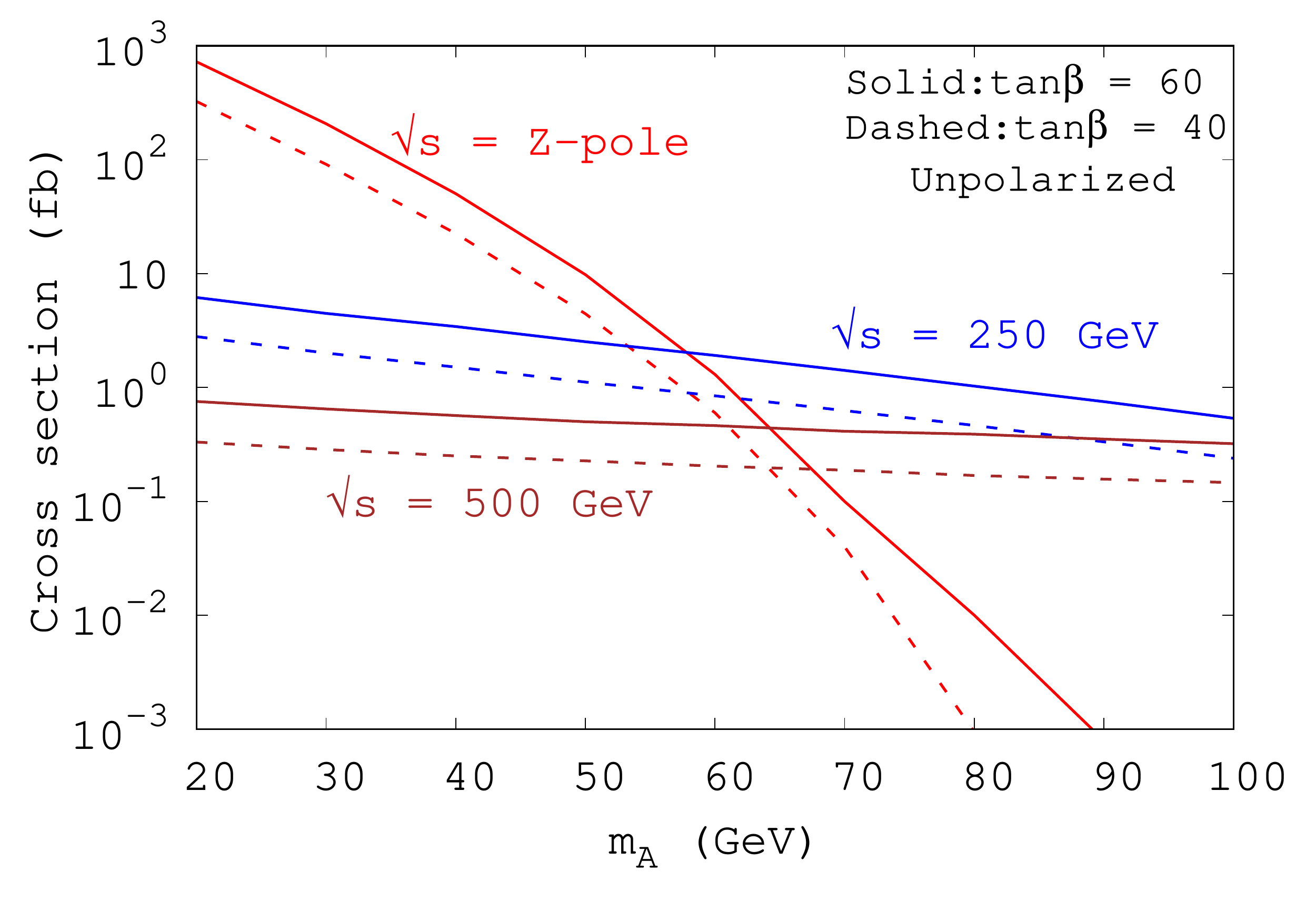}
 \caption{Production cross section of $e^+ e^- \to Z^*\gamma^* \to \tau\tau A$ as a function of the light boson mass at different center-of-mass energy. }
 \label{fig:x-section}
\end{center}
\end{figure}

The Yukawa process under the consideration is, 
$$e^+ e^- \to Z^*/\gamma^* \to \tau^+ \tau^- A \to  4 \tau.$$ 
Production cross section for  this channel in terms of $m_A$ for various  center-of-mass energy is depicted in 
Fig.~\ref{fig:x-section}. The cross-section is produced with unpolarized beams and will increase for polarized beam. 
For our analysis  we have polarized beam with $P(e^+,e^-)=(+30\%,-80\%)$~\cite{Behnke:2013lya}. 
Since the $A\tau\tau$ Yukawa coupling is proportional to $\tan\beta$, cross-section increases as $\tan\beta$ increases. 
Although  it is easier to produce a light $A$ at $Z$-pole, the taus coming from the decay of $A$ will have small momenta and 
 will remain undetected. The  250 GeV center-of-mass energy is perfect to explore the leptophilic Yukawa structure.
The four tau leptons eventually decay either hadronically (65\%) or leptonically (35\%). The signal events are identified  as 
$$3 \ j_\tau + X, \hspace{1cm} X \equiv j_\tau\ /j \ /\ell_\tau,$$ where $j_\tau$ is a $\tau$-tagged jet; $j$ is an  untagged jet, whereas 
$\ell_\tau \equiv e/\mu$ is leptons from the decay of $\tau$. The total number of objects is four. The inclusion of a lepton 
in the final state helps to increase signal events since leptonic decay of a tau lepton is substantially large.
 
 The dominant background to this channel comes from the $e^+e^- \to Z Z \to 4\tau$ and $e^+e^- \to Z Z \to 2\tau \ 2 j$ 
 processes where mis-identification of light jet into a $\tau$-tagged jet mimics the signal in the latter case. There are subdominant background 
 coming from the $e^+e^- \to Z h$ process and we have also considered it. Parton level production cross-section of  the 
 $4~\tau$ background process is $\sim 6.6 fb$  whereas the cross-section for the $2\tau\ 2j$ process is 
 $\sim 250 fb$ at 250 GeV ILC with polarized beam. 
 
\subsection{Event simulation and selection}

 Parton level signal and background events are simulated with $\texttt{MadGraph5\_aMC@NLO}$ \cite{Alwall:2011uj,Alwall:2014hca} and \texttt{PYTHIA8} \cite{Sjostrand:2006za,Sjostrand:2014zea} was used for the subsequent decay, showering and hadronization. The $\tau$ decays 
are incorporated via \texttt{TAUOLA} \cite{Jadach:1993hs} integrated in $\texttt{MadGraph5\_aMC@NLO}$. We have used \texttt{Delphes3} \cite{deFavereau:2013fsa} with the ILD detector card to simulate the detector effects. Jets are clustered using the anti-kT algorithm\footnote{In general, at a linear collider jets are clustered using the Durham algorithm. However recently it has been showed that anti-kt or the newly proposed Valencia algorithm can also be used for lepton collider analysis~\cite{Boronat:2014hva}.} 
\cite{Cacciari:2008gp} with $R = 0.4$. In \texttt{Delphes3},  we assumed the tagging efficiency of 
$\tau$ jets is 60\% and the corresponding mis-tagging rate is 0.5\% in accordance with tau tagging efficiency at LHC with multivariate analysis~\cite{CMS-PAS-TAU-16-002}. ILC being a lepton collider is expected to have same or better tagging efficiency for jets due to improved track momentum and jet energy resolution and our results are conservative. We have also used the Delphes jet charge measurement to make opposite sign jet pair.
 
We imposed the \textit{pre-selection criteria} that all the jets and leptons should have minimum energy of 20 GeV and should have $|\eta| <2.3$ which corresponds to $|\cos\theta| < 0.98$. Using the selected events we then move on to reconstruct the parent $\tau$-leptons.

\subsection{Collinear approximation and reconstruction of A}
At a lepton collider all the four components of initial and final state energy-momentum is known and it is possible to reconstruct 
 four taus by using the collinear approximation which assumes that the missing energy from the decay of tau lepton is 
collinear to the visible part of the decay. This approximation is true when tau lepton is boosted enough and in the Yukawa process discussed here, one of the distinct feature is the hard energy spectrum\footnote{This was used to search for Yukawa production ($b\bar{b}A/h \to 2b2\tau$) in three jet final state for Type-II 2HDM at LEP by OPAL Collaboration~\cite{Abbiendi:2001kp}.} of the pseudoscalar which ensures the applicability of this approximation.

The energy momentum conservation equations are,
\beas
\vec{p}(\tau_1)+\vec{p}(\tau_2)+\vec{p}(\tau_3)+\vec{p}(\tau_4) &=& \vec{0},\\
E(\tau_1)+E(\tau_2)+E(\tau_3)+E(\tau_4) &=& \sqrt{s}.
\eeas
 
 Now let us assume that the visible 4-momentum of the $i$-th object from $\tau$ decay takes $z_i$ amount of the original momenta $i.e.$ $p^\mu(j_i) = z_i  \ p^\mu (\tau_i)$ where $j_i$ is either a $\tau$-tagged jet or a light jet or a lepton. We use the visible four momentum and solve the above set of equations for $z_i$. The physical  solutions should yield that $0 < z_i < 1$. However, due to finite momentum resolution of the jets and since we are dealing with at least 3 $\tau$-tagged jets in the final state there will be ample uncertainty in the solution of collinear approximation. To accommodate this, we 
 have relaxed the condition on $z_i$ that it can go up to 1.1~\cite{Kanemura:2011kx}. Now using the $z_i$ we can reconstruct the momentum of the tau-leptons and finally reconstruct the pseudoscalar. 
 
 Since there are four $\tau$s, there will be four possible opposite sign tau-pair combination and one of them is coming from $A$. To identify the $A$ resonance without any ambiguity we use the following method :

\begin{figure}[t!]
\begin{center}
 \includegraphics[width=6.8cm]{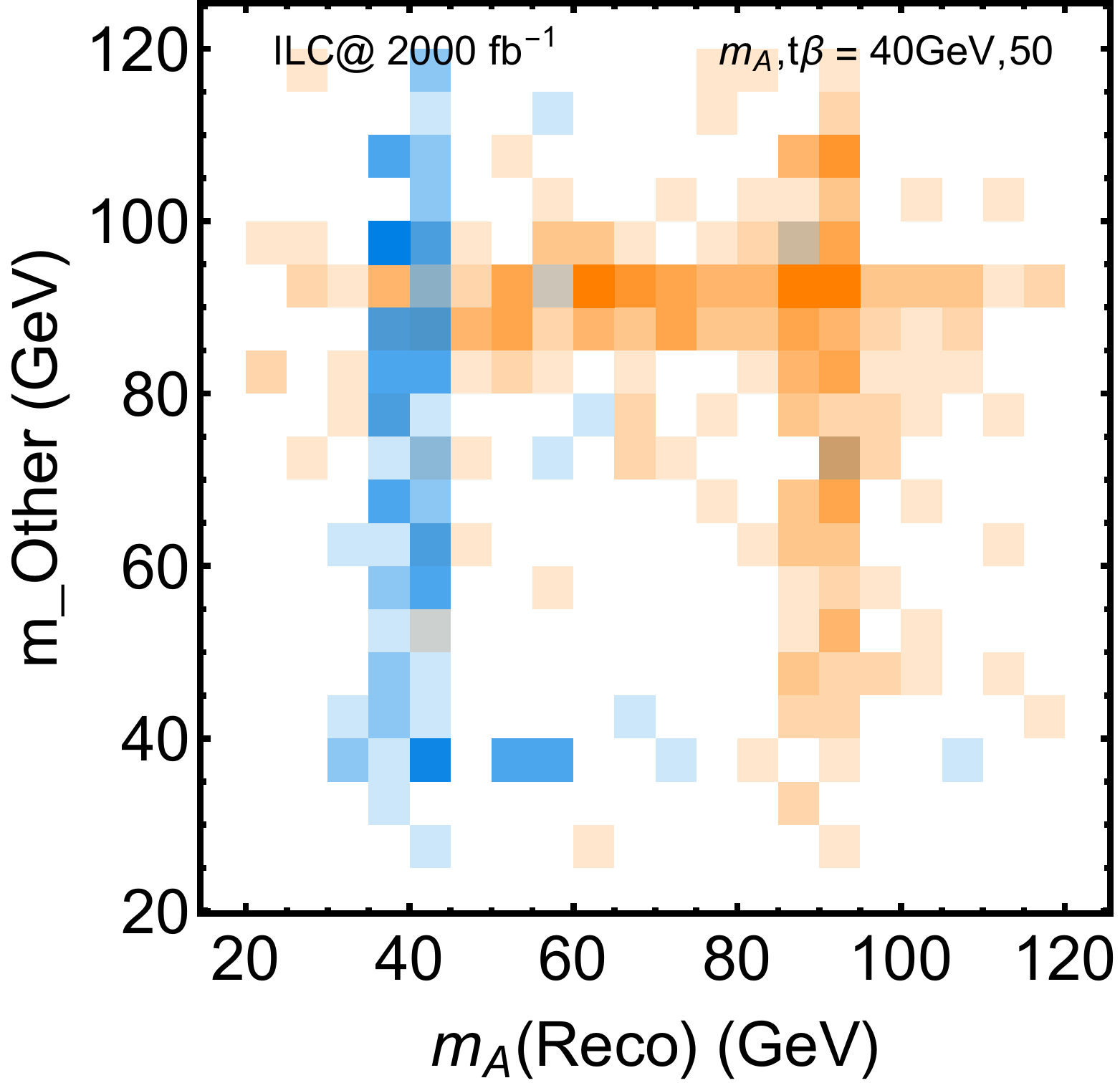}\hskip15pt
 \includegraphics[width=8.2cm]{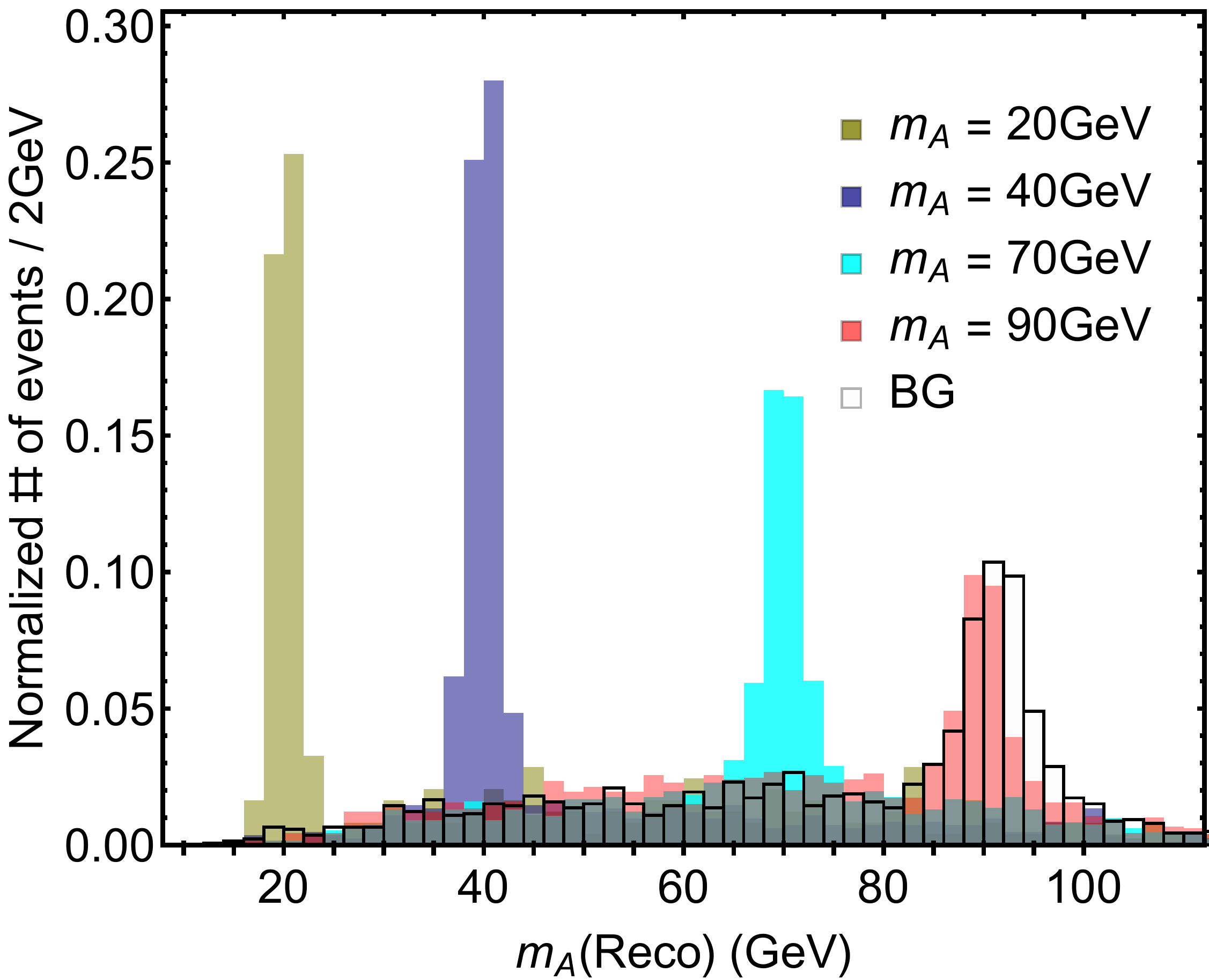}
\end{center}
\caption{\emph{Left Panel:} Density plot of $m_A(Reco)$ and $m_{\rm{Other}}$ for signal benchmark point ($m_A = 40$ GeV and $\tan\beta =50$) in blue and for background events in orange. See text for definitions of invariant masses. Signal and background events are 
generated at $\sqrt{s}=$250 GeV with ILC environment with integrated luminosity of 2000 $fb^{-1}$. \emph{Right panel:} Normalized invariant mass distribution of the reconstructed pseudoscalar using collinear approximation for different pseudoscalar mass.}
\label{fig:inv_mass_all}
\end{figure}

 \begin{itemize}
 \item \emph{The highest energy $\tau$ out of the four is unlikely to come from the pseudoscalar since the maximum available energy for $A$ is 125 GeV($\sqrt{s}/2$), whereas energy of highest $\tau$ can be 125 GeV also. Hence is it reasonable to assume that the highest energy tau is coming from the decay of $Z$ and did not radiated an $A$.}
 \item From the remaining 3 taus there are two possible opposite sign combinations. 
 \item Among the two possible combinations we choose the combination which gives highest transverse momentum($p_T$) since they are likely to come from the decay of $A$. The invariant mass calculated from this combination is denoted as $m_A(Reco)$ and the distribution of it should display the resonance peak.  The  invariant mass from the other opposite sign tau pair is denoted as $m\_{Other}$. 
\end{itemize}

To show the efficacy of the method we have plotted  both the invariant mass distribution ($m_A(Reco) \ \& \ m\_{Other}$) in the left panel of Fig.~\ref{fig:inv_mass_all} for a benchmark pseudoscalar mass of 40 GeV with $\tan\beta=50$. The signal events are displayed in blue and the background events are shown in orange. All the events are generated at 250 GeV ILC with integrated luminosity 
amount to 2000 $fb^{-1}$. Evidently the $m_A(Reco)$ clustered around 40 GeV whereas $m\_{Other}$ is arbitrary. 
The background events are clustered near the $Z$-boson mass as the dominant background is from $ZZ$. 
In the right panel of Fig.~\ref{fig:inv_mass_all} we have plotted the reconstructed invariant mass distribution $m_A(Reco)$ for different values of $m_A$ and the background events . As $m_A$ increases the invariant mass peak becomes broader as the 
decay with of $A$ is proportional to its mass.   It is evident that the method described before can be used for success full mass reconstruction.

\section{Results}\label{sec:result}

\begin{table}
\begin{center}
\begin{tabular}{|c||c|c|c||c|}
\hline
 \multicolumn{5}{|c|}{\bf Pre-selection cut : Energy $>$ 20 GeV. $|\eta| <$ 2.3 }\\
\hline
$\mathcal{L}$ = 2000  $fb^{-1}$ & \multirow{2}{*}{Signal} & \multicolumn{2}{c|}{Background}   & \multirow{2}{*}{Significance}   \\ \cline{1-1} \cline{3-4} 
                               &                         & 4$\tau$ & 2$\tau$ \ 2 j &                \\ \hline
Pre-selection cut               & 106 [100\%]            & 242 [100\%]       & 98[100\%]            & 5.5 \\ \hline
Collinear approx  &&&&\\
$0 < z_i <1.1$               & 91 [86.0\%]           & 217[89.7\%]      & 69[70.4\%]         & 5.1 \\ \hline
$m_A\pm10$GeV                  & 66 [62.3\%]           & 32   [14.9\%]       & 10[10.2\%]            & 8.5  \\ \hline
\end{tabular}
\caption{Cut flow for $m_A = 40$~GeV and $\tan \beta=50$ with integrated luminosity of 2000 $fb^{-1}$. } 
\label{tab:cut-fow}
\end{center}
\end{table}
\begin{figure}[!t]
\begin{center}
 \includegraphics[width=7.5cm]{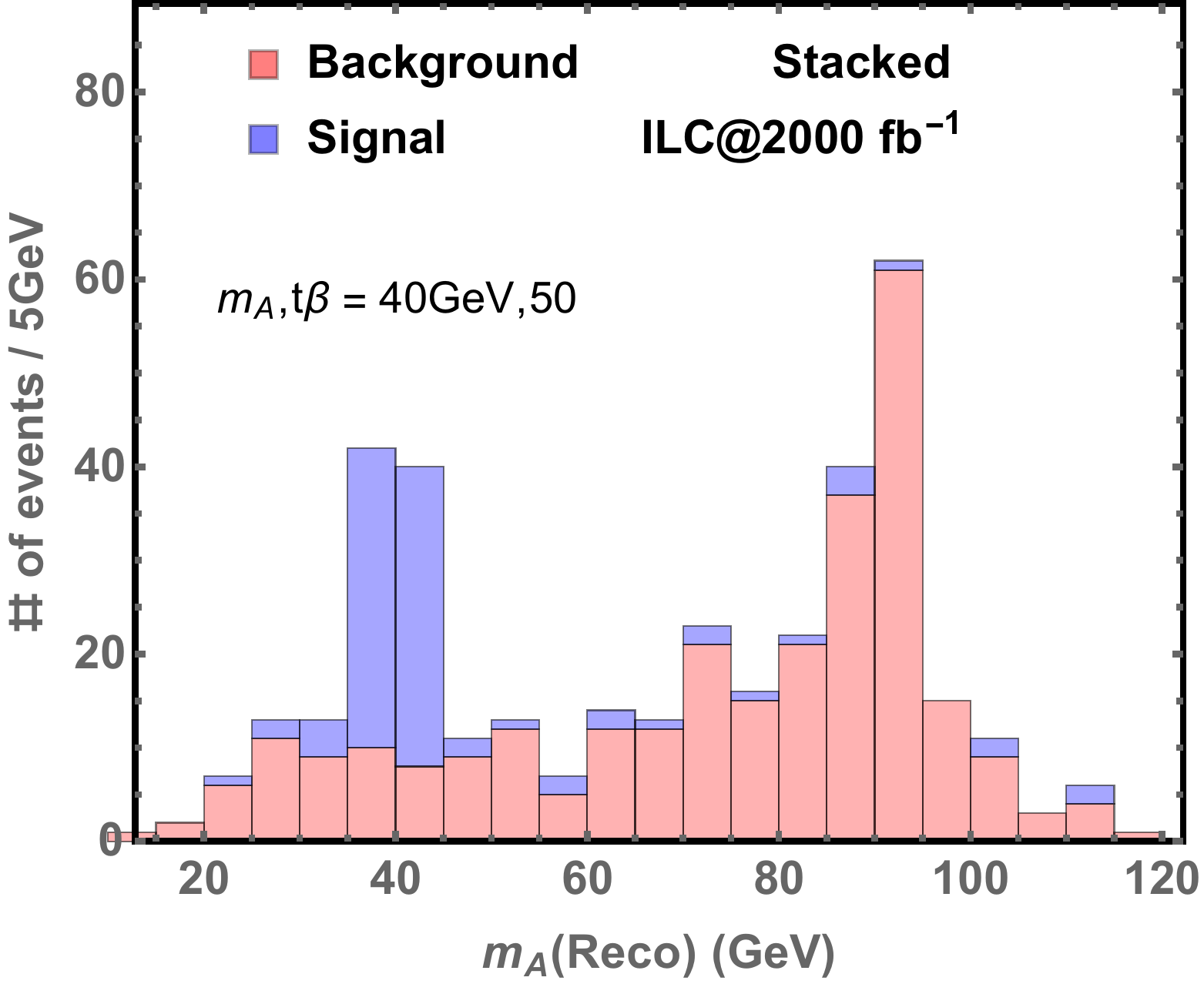}
  \includegraphics[width=7.5cm]{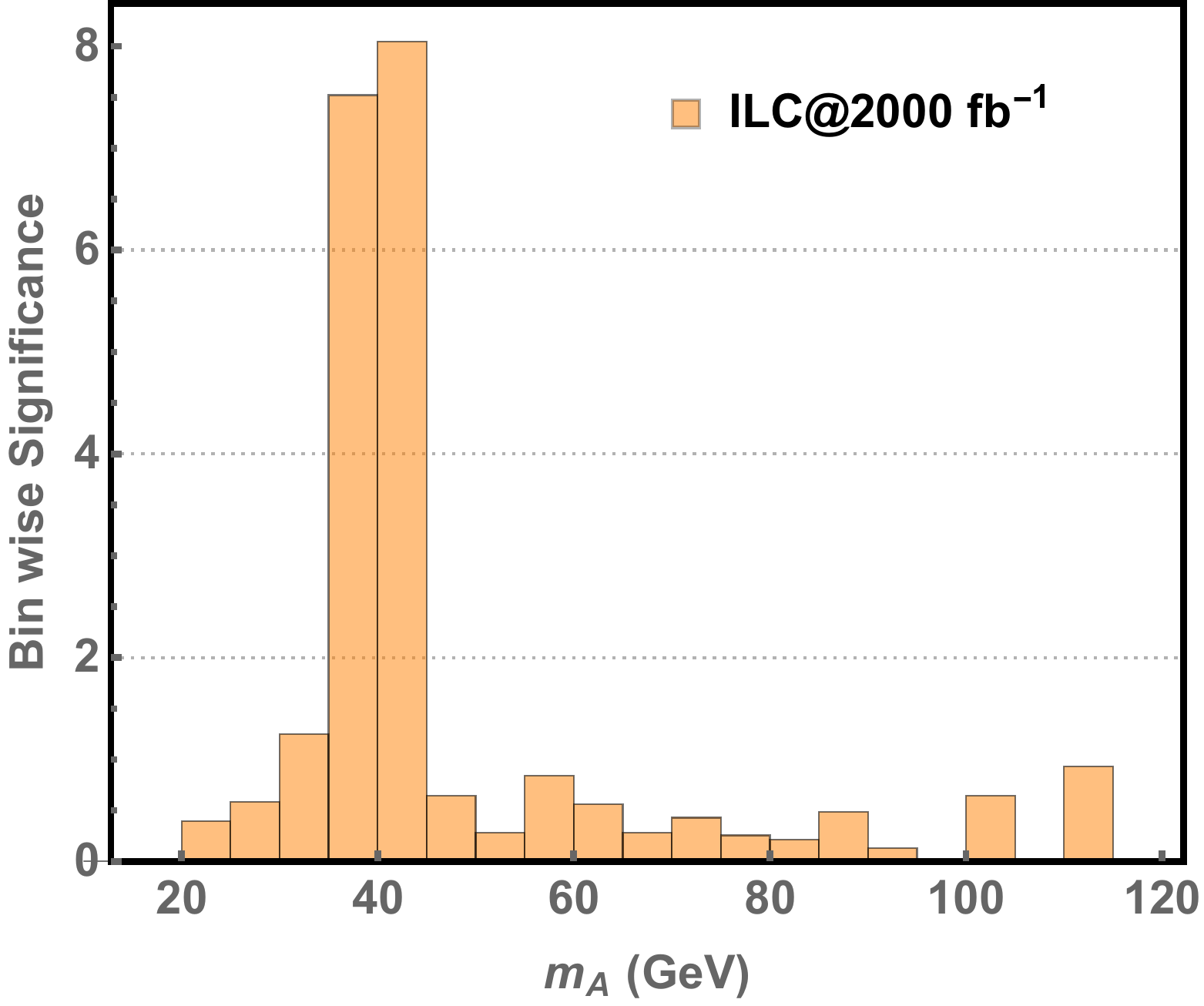}
\end{center}
\caption{Signal and background events at ILC250 at 2000 $fb^{-1}$ are plotted in the left panel whereas the binwise significance 
is plotted in the right panel. Expectedly the bin wise significance peaks only at the pseudoscalar mass.}
\label{fig:benchamrk}
\end{figure}

We have shown in the previous section that the collinear approximation can be used to reconstruct the mass and we will use  the 
reconstructed invariant mass to further minimize the background events. 
 The cut flow table for the benchmark signal ($m_A = 40$ GeV and $\tan\beta =50$) and background events are shown in 
Table.~\ref{tab:cut-fow} with integrated luminosity of 2000 $fb^{-1}$. Since the background cross-section is low it is possible to 
achieve $5~\sigma$ significance at the pre-selection level and when we use the invariant mass window cut the significance increased 
to more than  $8~\sigma$. Here signal significance has been calculated using the following expression,
\begin{equation}
\mathcal{S} = \sqrt{2\left[(S+B)\textrm{ln}\left(1+\frac{S}{B}\right)-S\right]}, 
\end{equation}
 where $S(B)$ are number of signal (background) events after the cuts. In the left panel of Fig.~\ref{fig:benchamrk} 
 we display the reconstructed $m_A$ distribution for the signal and background events where both the events are stacked together. 
 Here we have used the same signal benchmark ($m_A = 40$ GeV and $\tan\beta =50$). The number of events are 
 computed for ILC with integrated luminosity of 2000 $fb^{-1}$. Also we have plotted the binwise signal significance in the 
 right panel of Fig.~\ref{fig:benchamrk}. As expected, the background events are clustered near the $Z$-boson mass in Fig.~\ref{fig:benchamrk} 
 (left panel) and the binwise significance is large ($\sim8~\sigma$) near true $m_A$ value. 

\begin{figure}[!t]
\begin{center}
    \includegraphics[width=8cm]{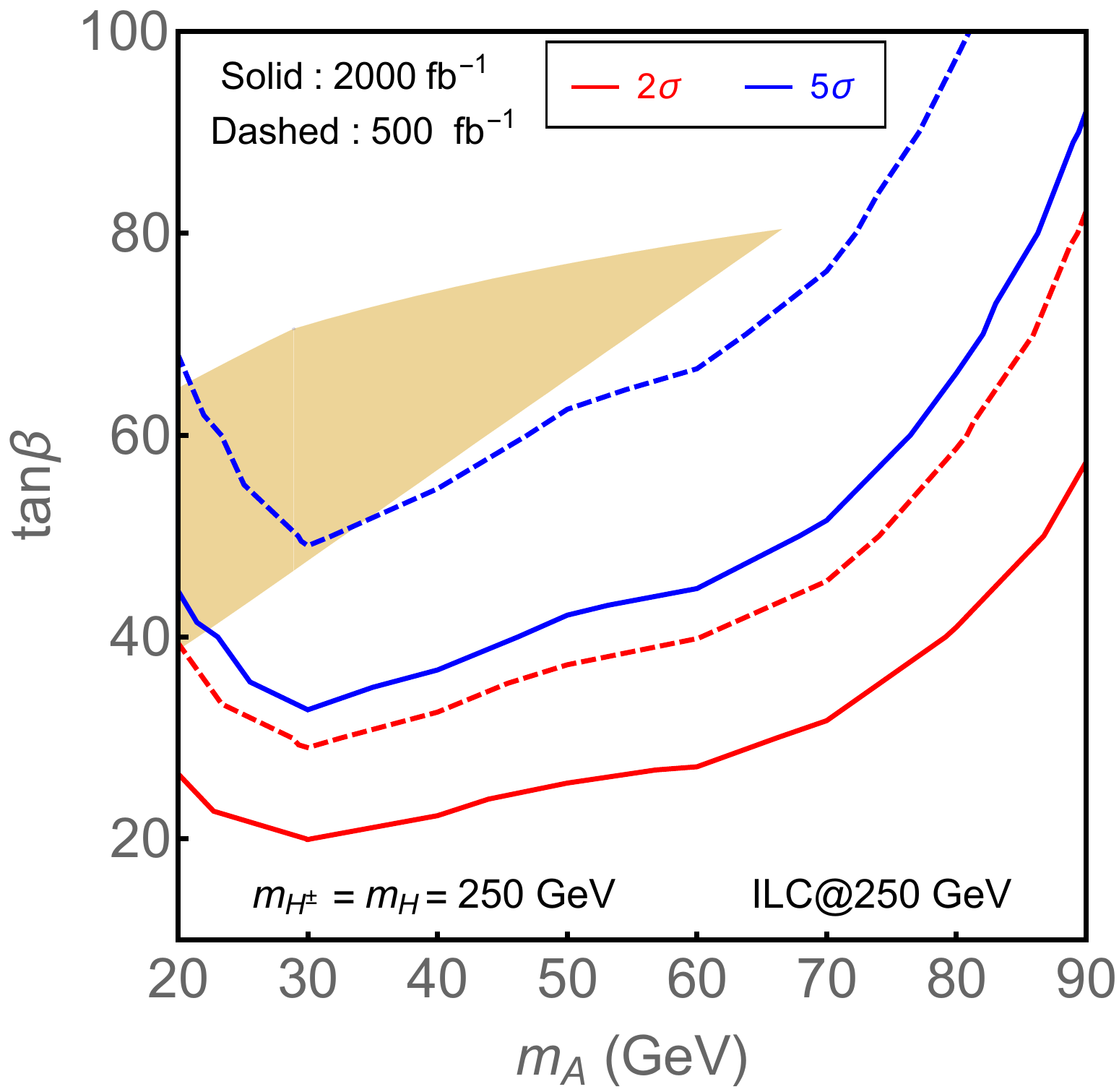}
\end{center}
\caption{Reach of the 250 GeV $e^+e^-$ collider in the $m_A$--$\tan\beta$ parameter space of the Type-X 2HDM. 
The significance increases for larger $\tan\beta$ as the 
signal production cross section is almost proportional to $\tan^2\beta$. The light yellow region 
can explain the $(g-2)_\mu$ anomaly at $2~\sigma$ after applying the lepton universality constraints.
}
\label{fig:param-scan}
\end{figure}

Equipped with the mass reconstruction method we have scanned full the $m_A - \tan\beta$ parameter space and computed the 
signal significance at ILC with $\sqrt{s} =$ 250 GeV with integrated luminosity of 500 $fb^{-1}$ and 2000~$fb^{-1}$. 
The $2~\sigma$ exclusion and $5~\sigma$ discovery contours are shown in Fig.~\ref{fig:param-scan}. 
As $m_A$ decreases the decay products become soft which leads to weak bound and at higher $m_A$ signal cross-section decreases and 
also as we move towards the $Z$-boson mass it it very difficult to distinguish the signal and background events. 
Consequently the bound becomes weak at higher $m_A$ values. We have also shown the allowed parameter space which can explain 
$(g-2)_\mu$ after satisfying the lepton universality constraints coming from the tau decay and $Z\to \ell\ell$ 
measurements~\cite{ALEPH:2005ab,Amhis:2016xyh}.  Our result is obtained following the analysis in \cite{Chun:2016hzs} and the update in \cite{Chun:2019oix}.

It is evident that a large portion of the parameter space which is 
favoured by the muon anomaly can be thoroughly scrutinized at ILC250 even with 500 $fb^{-1}$ luminosity where the exclusion 
limit goes below $\tan\beta =$ 40. With higher luminosity even more parameter can be explored for this model.

\section{Conclusion}\label{sec:conclusion}

A light pseudoscalar in Type-X 2HDM at large $\tan\beta$ can explain the observed deviation of the muon anomalous magnetic moment and thus it is worthwhile to test the scenario at hadron or lepton colliders like LHC or ILC. 
Due to the hadrophobic nature of the pseudoscalar it is very hard to look for it at LHC unless the heavier Higgs bosons,
$H^\pm$ and $H$, are lighter than about 200 GeV. On the other hand, future lepton colliders appear ideal to probe the relevant parameter  space through  the (tau) Yukawa process independent of the heavy Higgs masses.

We demonstrated that it is possible to utilize the \emph{Higgs-factory}, e.g., ILC at 250 GeV, for testing the model  regardless of the specific values of heavier Higgs masses.
A realistic analysis with the $4 \tau$ final states is presented to  reconstruct the light pseudoscalar by using the collinear approximation.
 The entire parameter space explaining the muon $(g-2)$ anomaly is shown to be explored 
at $5~\sigma$ with  integrated luminosity of 2000 $fb^{-1}$.

\section{Appendix}

The collinear approximation works better for larger mass of the pseudoscalar. 
As $m_A$  increases the momentum of its decay products increases and the taus are more boosted, which is essential for this approximation. 
This is shown in Table~\ref{tab:coll}. However the invariant mass peak becomes broader for larger mass, and the window 
of 20 GeV does not contain the full resonance which results a decrease in accepted number of events after invariant mass cut. 
Also when $m_A$ is very close to $90$ GeV our assumption that the highest energy jet is coming from $Z^*/\gamma^*$ falls apart 
which also worsens the situation. We have checked that the assumption holds good up to 80 GeV which 
is already beyond the range of our interest.

\begin{table}[!t]
\begin{center}
\begin{tabular}{|c|c|c|c|}
\hline
$m_A$ (GeV) & Pre-selection    & After Collinear & After $m_A\pm10$ GeV \\ \hline
20          & 336   & 245(72.9\%)     & 132(39.3\%)          \\ \hline
30          & 790   & 657(83.2/\%)    & 472(59.7\%)          \\ \hline
40          & 960   & 826(76/\%)      & 597(62.2/\%)          \\ \hline
50          & 1078  & 954(88.5/\%)    & 634(58.8/\%)         \\ \hline
60          & 1301  & 1166(89.6/\%)   & 725(55.7/\%)         \\ \hline
70          & 1512  & 1353(89.5\%)    & 801(53.0/\%)       \\ \hline
80          & 1694  & 1540(90.9/\%)   & 798(47.1/\%)         \\ \hline
90          & 1978  & 1803(91.2/\%)   & 713(36.0/\%)          \\ \hline
\end{tabular}
\caption{Energy of jet and lepton $>$ 20 GeV \& $|\eta| <$ 2.3. Efficiency at different mass of A.
As mass increases collinear approximation becomes better since high $m_A$ equivalent to more collinear tau decay.
The mass reconstruction becomes poor since we are taking a small invariant mass window. }
\label{tab:coll}
\end{center}
\end{table}

\providecommand{\href}[2]{#2}\begingroup\raggedright\endgroup

\end{document}